\documentclass[12pt]{article}
\usepackage{epsfig}
\begin{document}
\begin{center}
{\Large \bf Higher-order Threshold Corrections for Single Top Quark Production}
\end{center}
\vspace{2mm}
\begin{center}
{\large Nikolaos Kidonakis\footnote{Presented at DIS 2007, Munich, Germany, 
April 16-20, 2007}}\\
\vspace{2mm}
{\it Kennesaw State University, Physics \#1202\\
1000 Chastain Rd., Kennesaw, GA 30144-5591}\\
\end{center}
 
\begin{abstract}
I discuss single top quark production at the Tevatron and the LHC. 
The cross section, including soft-gluon threshold corrections through 
NNNLO, is presented for each partonic channel. The higher-order corrections
provide significant contributions to the single top cross sections at 
both colliders.
\end{abstract}

\section{Introduction}

Single top quark production at hadron colliders can proceed through 
three distinct partonic processes: the $t$ channel 
($qb \rightarrow q' t$ and ${\bar q} b \rightarrow {\bar q}' t$)
which involves the exchange of a spacelike $W$ boson, 
the $s$ channel ($q{\bar q}' \rightarrow {\bar b} t$) which proceeds
via a timelike $W$ boson, and 
associated $tW$ production ($bg \rightarrow tW^-$)~\cite{url}.
The $t$ channel processes are numerically the largest at both the 
Tevatron and the LHC. At the Tevatron the $s$ channel is second in 
magnitude and $tW$ production has the smallest cross section. At the
LHC, $tW$ production has a much bigger cross section than the $s$ channel.

The cross sections for all these processes receive contributions 
from soft-gluon emission which can be dominant near 
threshold~\cite{NKtev,NKlhc}.
Threshold resummation organizes these contributions and can be used 
to compute higher-order corrections for many 
processes~\cite{NNLO,NNNLO,EW}. 
For the partonic process with momenta $p_1 +p_2 \rightarrow p_3+p_4$
we define $s=(p_1+p_2)^2$, $t=(p_1-p_3)^2$, $u=(p_2-p_3)^2$
and $s_4=s+t+u-m_3^2-m_4^2$. Near threshold, $s_4$ approaches zero
and the soft-gluon corrections take the form
$[\ln^l(s_4/m_t^2)/s_4]_+$, where $m_t$ is the top quark mass and 
$l \le 2n-1$ for the $n$-th order corrections. 
We calculate these corrections through next-to-next-to-next-to-leading order 
(NNNLO) in the strong coupling $\alpha_s$ at next-to-leading logarithmic 
(NLL) accuracy at the 
Tevatron~\cite{NKtev} and the LHC~\cite{NKlhc}. This 
requires one-loop calculations in the eikonal approximation.
  
The NLO soft-gluon corrections can be written in the form  
\begin{eqnarray}
\frac{d^2{\hat\sigma}^{(1)}}{dt \, du}
=F^B \frac{\alpha_s(\mu_R^2)}{\pi} \left\{
c_{3} \left[\frac{\ln(s_4/m_t^2)}{s_4}\right]_+
+c_{2} \left[\frac{1}{s_4}\right]_+
+c_{1}^{\mu} \, \delta(s_4)\right\}
\nonumber
\end{eqnarray}
where $F^B$ is the Born term for each channel and $\mu_R$ is the 
renormalization scale.
For the $t$ and $s$ channels the leading logarithm coefficient is 
$c_{3}^{t,s}=3C_F$ while for the $tW$ channel it is $c_3^{tW}=2(C_F+C_A)$, 
where $C_F=(N_c^2-1)/(2N_c)$ and $C_A=N_c$ with $N_c=3$ the number of colors.
The NLL coefficient is
$c_2^{s}=-\frac{7}{4}C_F+2C_F\ln\left(\frac{s(s-m_t^2)}{(t-m^2_t)(u-m^2_t)}
\right)-2C_F \ln\left(\frac{\mu_F^2}{m_t^2}\right)$ 
for the $s$ channel, where $\mu_F$ is the 
factorization scale, and similar expressions can be given for the other 
channels. The complete virtual corrections ($\delta(s_4)$ terms)
cannot be derived from threshold resummation but one can derive the 
factorization and renormalization scale terms denoted by $c_1^{\mu}$ 
in the above equation~\cite{NKtev}.

The NNLO soft-gluon corrections for the $t$ and $s$ channels are 
\begin{eqnarray}
\frac{d^2{\hat\sigma}^{(2)}_{t,s}}{dt \, du}
&=&F^B \frac{\alpha_s^2(\mu_R^2)}{\pi^2}
\left\{\frac{1}{2} c_3^2
\left[\frac{\ln^3(s_4/m_t^2)}{s_4}\right]_+ \right.
\nonumber \\ && \hspace{20mm} \left.
{}+\left[\frac{3}{2} c_3 \, c_2
-\frac{\beta_0}{4} c_3 +C_F \frac{\beta_0}{8}\right]
\left[\frac{\ln^2(s_4/m_t^2)}{s_4}\right]_+ \right\}
\nonumber
\end{eqnarray}
plus subleading terms~\cite{NKtev}, where the appropriate expression 
for $F_B$ and $c_3$, $c_2$ for each channel must be used, and where 
$\beta_0=(11C_A-2n_f)/3$ is the lowest-order $\beta$ function, with $n_f$ 
the number of light quark flavors.
A similar expression holds for the $tW$ channel (by deleting 
$C_F \beta_0/8$ above).
Since this is a NLL calculation, only the leading and NLL terms shown above are
complete. However we can also calculate exactly terms involving $\mu_F$ and
$\mu_R$ as well as terms with $\zeta$ constants in the subleading logarithms. 
Complete expressions and further details are provided in Ref.~\cite{NKtev}.  

The NNNLO soft-gluon corrections for each channel can be written as
\begin{eqnarray}
\frac{d^2{\hat\sigma}^{(3)}}{dt \, du}
&=&F^B \frac{\alpha_s^3(\mu_R^2)}{\pi^3}
\left\{\frac{1}{8} c_3^3
\left[\frac{\ln^5(s_4/m_t^2)}{s_4}\right]_+ \right.
\nonumber \\ && \hspace{20mm} \left.
{}+\left[\frac{5}{8} c_3^2 c_2
-\frac{5}{48} \beta_0 c_3 (2 c_3-C_F) \right]
\left[\frac{\ln^4(s_4/m_t^2)}{s_4}\right]_+ \right\}
\nonumber
\end{eqnarray}
plus subleading terms~\cite{NKtev}, again with the appropriate expression 
for $F_B$ and $c_3$, $c_2$.

\section{Single top quark production at the Tevatron} 

\begin{figure} 
\begin{center}
\includegraphics[width=6.6cm]{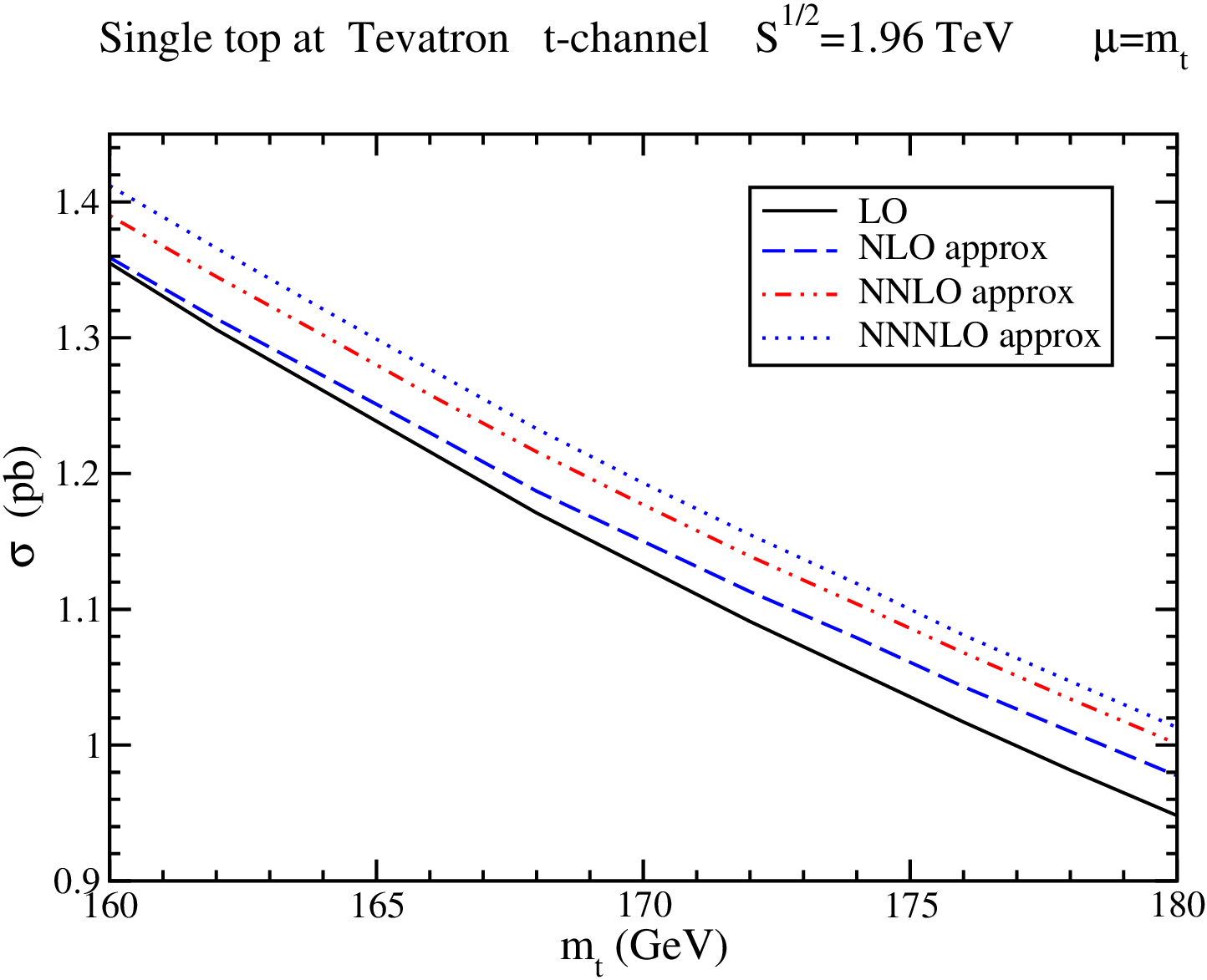}
\hspace{5mm}
\includegraphics[width=6.6cm]{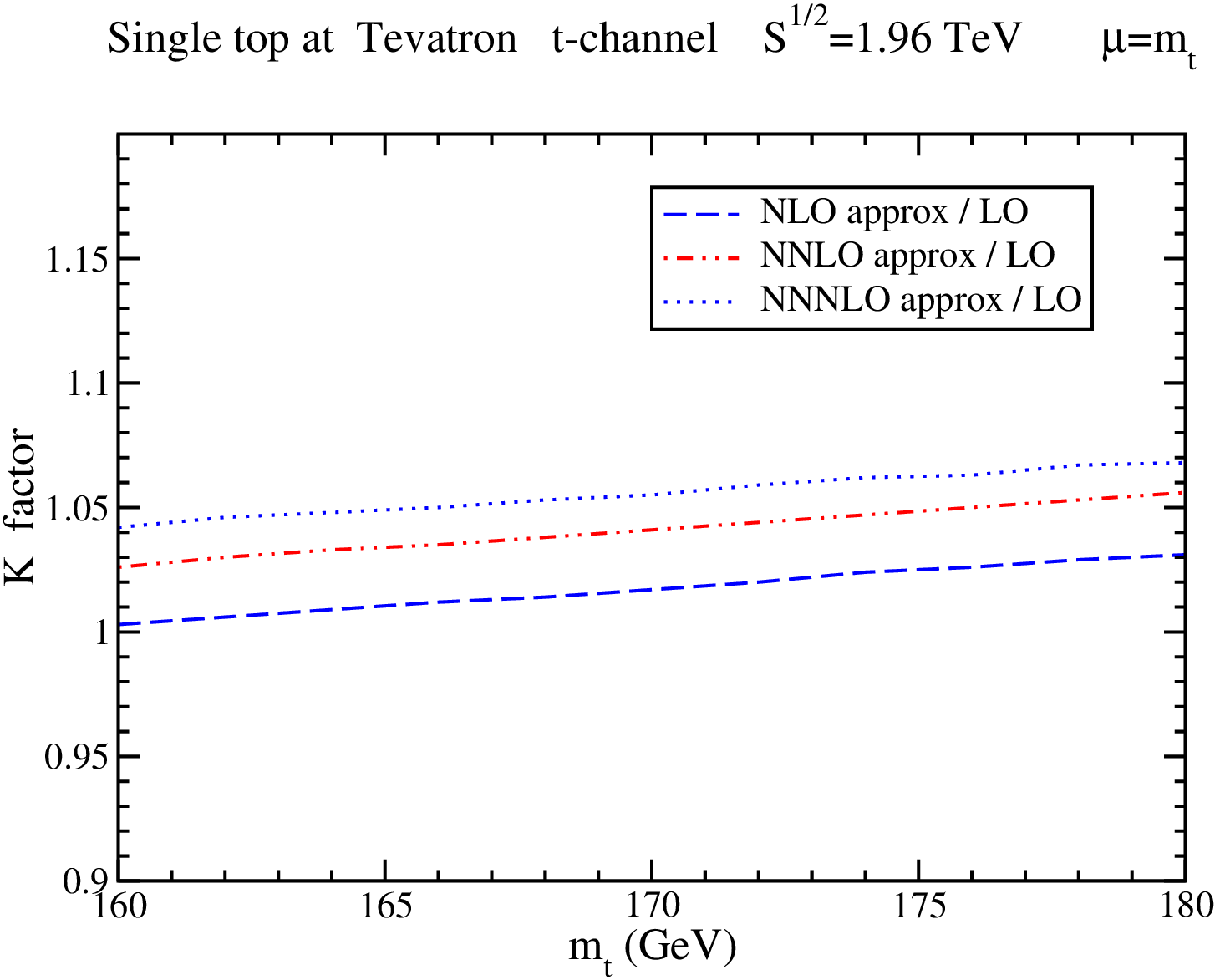}
\caption{$t$-channel single top quark cross section at the Tevatron.}
\end{center}
\label{Fig:t-chan.TeV}
\end{figure}

We now calculate the contribution of these corrections to the single top  
cross section at the Fermilab Tevatron. The MRST 2004 NNLO parton 
densities~\cite{MRST} are used for the numerical results. We find that the 
threshold corrections are dominant in all partonic channels. 

Figure 1 shows the results for the cross section in the $t$ channel. 
In the left-hand plot we show the leading-order (LO) cross section as well 
as the cross sections with the NLO, NNLO, and NNNLO soft-gluon corrections 
included versus the top quark mass $m_t$, with the factorization and 
renormalization scales set equal to $m_t$. On the right-hand plot we 
show the $K$ factors, which are the ratios of the higher-order cross sections 
to LO. We see that the corrections in this channel are relatively small.
Our best estimate for the cross section is calculated after matching to the 
exact NLO cross section~\cite{NLO}, i.e. by adding the soft-gluon corrections 
through NNNLO to the exact NLO cross section. Below we give results for 
two choices of the top quark mass, $m_t=170$ GeV and $m_t=175$ GeV. We find  
$\sigma^{t-{\rm channel}}(m_t=170 \,{\rm GeV})=1.17 \pm 0.06$ pb
and
$\sigma^{t-{\rm channel}}(m_t=175 \,{\rm GeV})=1.08 \pm 0.06$ pb.
The uncertainty indicated includes the scale dependence and the pdf 
uncertainties.

Figure 2 shows the results for the cross section and $K$ factors  
in the $s$ channel. In this channel the corrections are large, providing 
up to 65\% enhancement of the leading-order cross section.
After matching, we find
$\sigma^{s-{\rm channel}}(m_t=170 \, {\rm GeV})=0.56 \pm 0.03$ pb
and
$\sigma^{s-{\rm channel}}(m_t=175 \, {\rm GeV})=0.49 \pm 0.02$ pb.

The single top cross section at the Tevatron in the $tW$ channel is 
rather small, even though the $K$ factors are large (up to 85\% enhancement).
Our estimate for the cross section is  
$\sigma^{tW}(m_t=170 \, {\rm GeV})=0.15 \pm 0.03$ pb 
and
$\sigma^{tW}(m_t=175 \, {\rm GeV})=0.13 \pm 0.03$ pb.

For all three channels at the Tevatron the cross section for single anti-top 
production is identical to that shown above for single top production.

\begin{figure}
\begin{center}
\includegraphics[width=6.6cm]{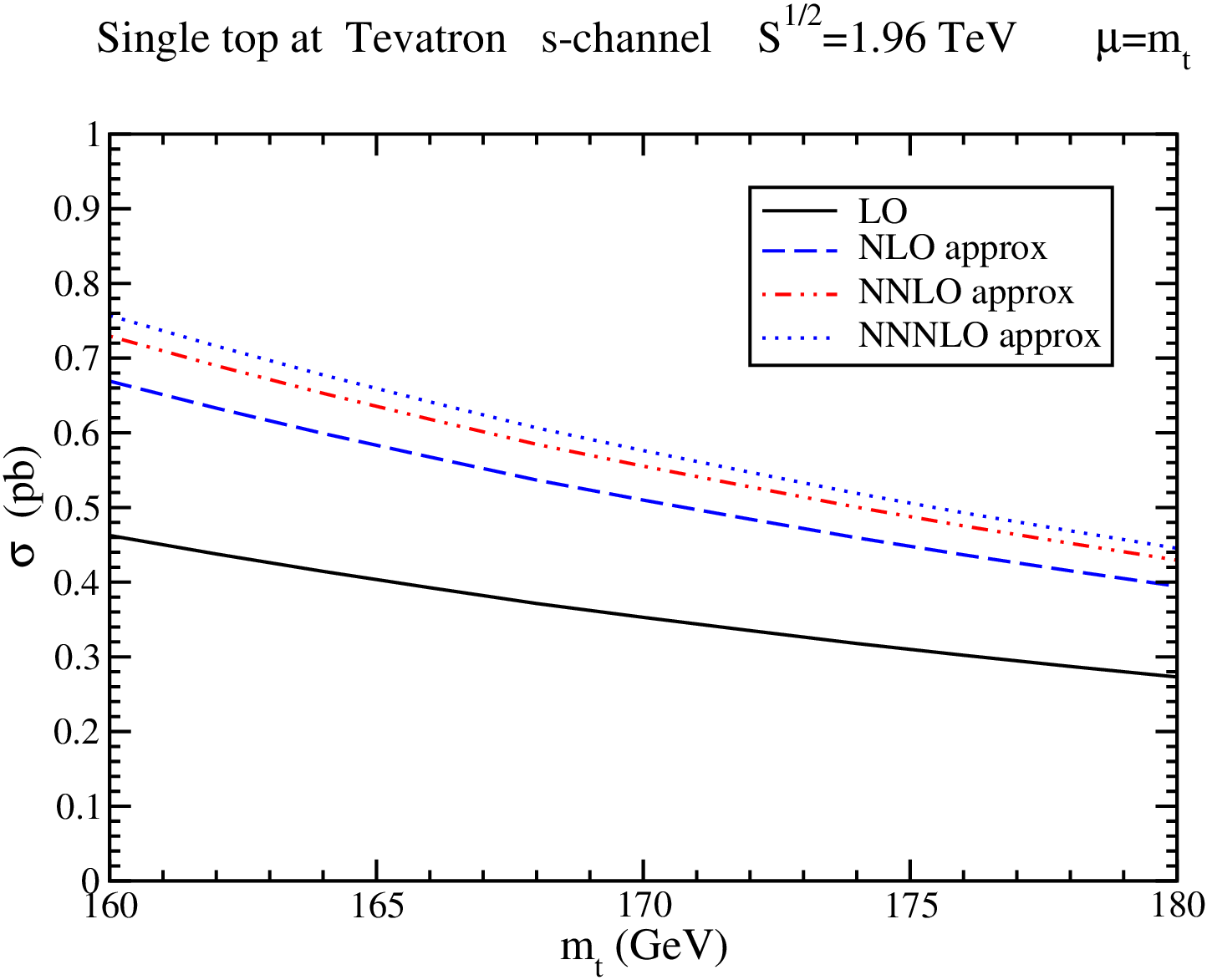}
\hspace{5mm}
\includegraphics[width=6.6cm]{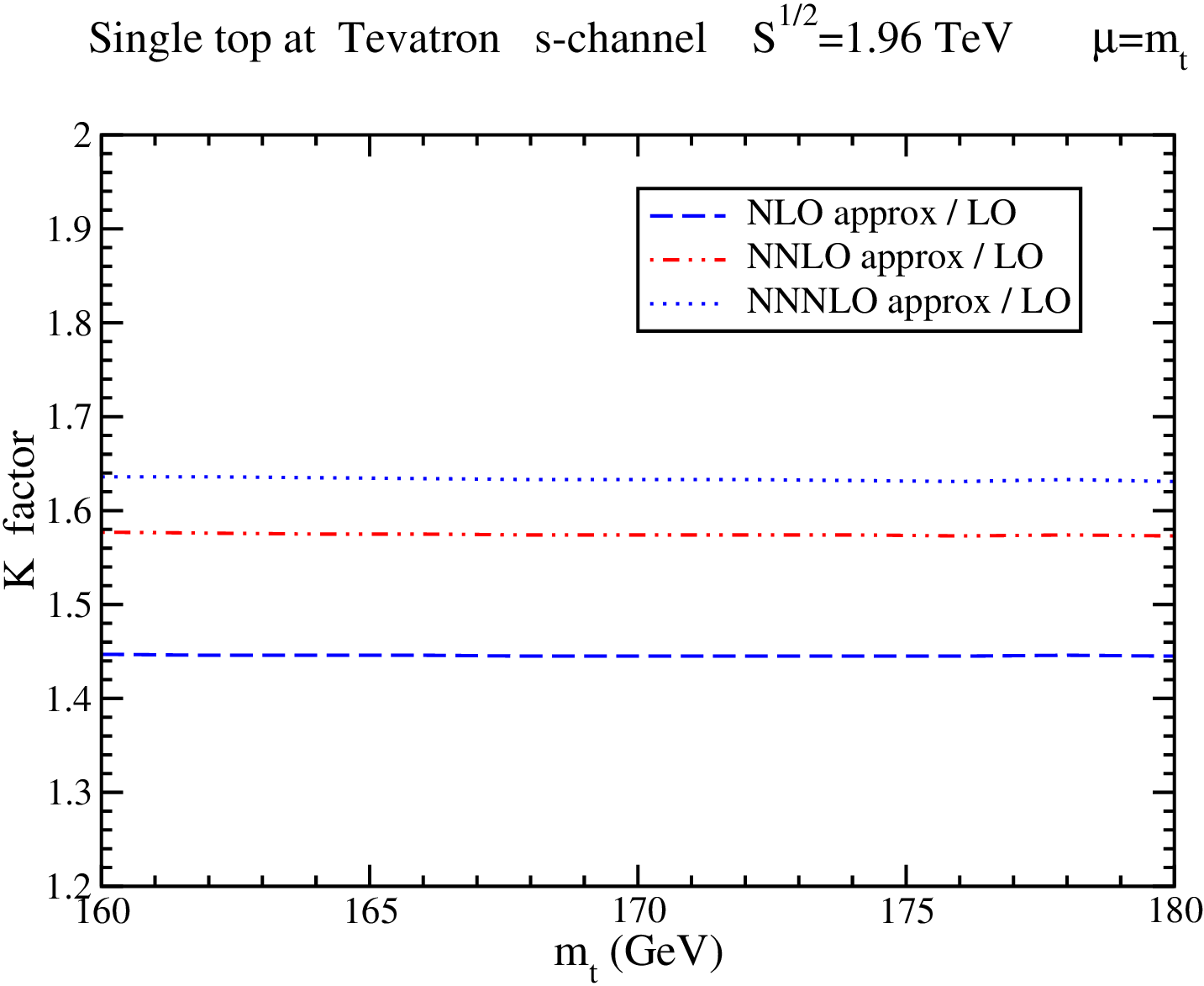}
\caption{$s$-channel single top quark cross section at the Tevatron.}
\end{center}
\label{Fig:s-chan.TeV}
\end{figure}

Finally, we note that there has been recent evidence for single top quark 
production at the Tevatron~\cite{D0} with a cross section consistent with 
the above results.

\section{Single top quark production at the LHC} 

Next we calculate the threshold corrections for the single top  
cross section at the CERN LHC.
It turns out that in the $t$ channel the threshold corrections are 
not a good approximation of full QCD corrections, hence we only update 
the exact NLO result \cite{NLO}, while in the $s$ and $tW$ channels the 
threshold approximation holds and we provide results including the NNNLO 
soft-gluon corrections. Also at the LHC the cross section for single top 
is different from that for single antitop production in the $t$ and 
$s$ channels.
 
The exact NLO cross section for single top production in the $t$ channel 
at the LHC is 
$\sigma^{t-{\rm channel}}_{\rm top} (m_t=170 \,{\rm GeV})=152 \pm 6$ pb 
and
$\sigma^{t-{\rm channel}}_{\rm top} (m_t=175 \,{\rm GeV})=146 \pm 5$ pb.
For single antitop production in the $t$ channel
the exact NLO cross section is
$\sigma^{t-{\rm channel}}_{\rm antitop} (m_t=170 \,{\rm GeV})=93 \pm 4$ pb
and
$\sigma^{t-{\rm channel}}_{\rm antitop} (m_t=175 \,{\rm GeV})=89 \pm 4$ pb.

Figure 3 (left) shows results for single top production in the $s$ channel 
at the LHC. The contribution from soft gluons is significant (up to 
55\% enhancement). 
After matching to the exact NLO cross section~\cite{NLO}, we find
$\sigma^{s-{\rm channel}}_{\rm top}(m_t=170\, {\rm GeV})=8.0^{+0.6}_{-0.5}$ pb
and
$\sigma^{s-{\rm channel}}_{\rm top}(m_t=175\, {\rm GeV})=7.2^{+0.6}_{-0.5}$ pb.

The corresponding results for single antitop production at the LHC in the 
$s$ channel are 
$\sigma^{s-{\rm channel}}_{\rm antitop}(m_t=170 \, {\rm GeV})=4.5 \pm 0.2$ pb
and 
$\sigma^{s-{\rm channel}}_{\rm antitop}(m_t=175 \, {\rm GeV})=4.0 \pm 0.2$ pb.
Here the soft-gluon corrections are somewhat smaller (less than 20\%).

Figure 3 (right) shows results for single top production at the LHC in 
the $tW$ channel. This channel has a significant cross section at the LHC. 
Also the soft-gluon corrections are quite large, providing around 60\% 
enhancement. 
After matching to the exact NLO cross section~\cite{Zhu}, we find
$\sigma^{tW}(m_t=170 \, {\rm GeV})=44 \pm 5$ pb
and
$\sigma^{tW}(m_t=175 \, {\rm GeV})=41 \pm 4$ pb.
The cross section for associated antitop production is identical to 
that for a top quark.

\begin{figure}
\begin{center}
\includegraphics[width=6.6cm]{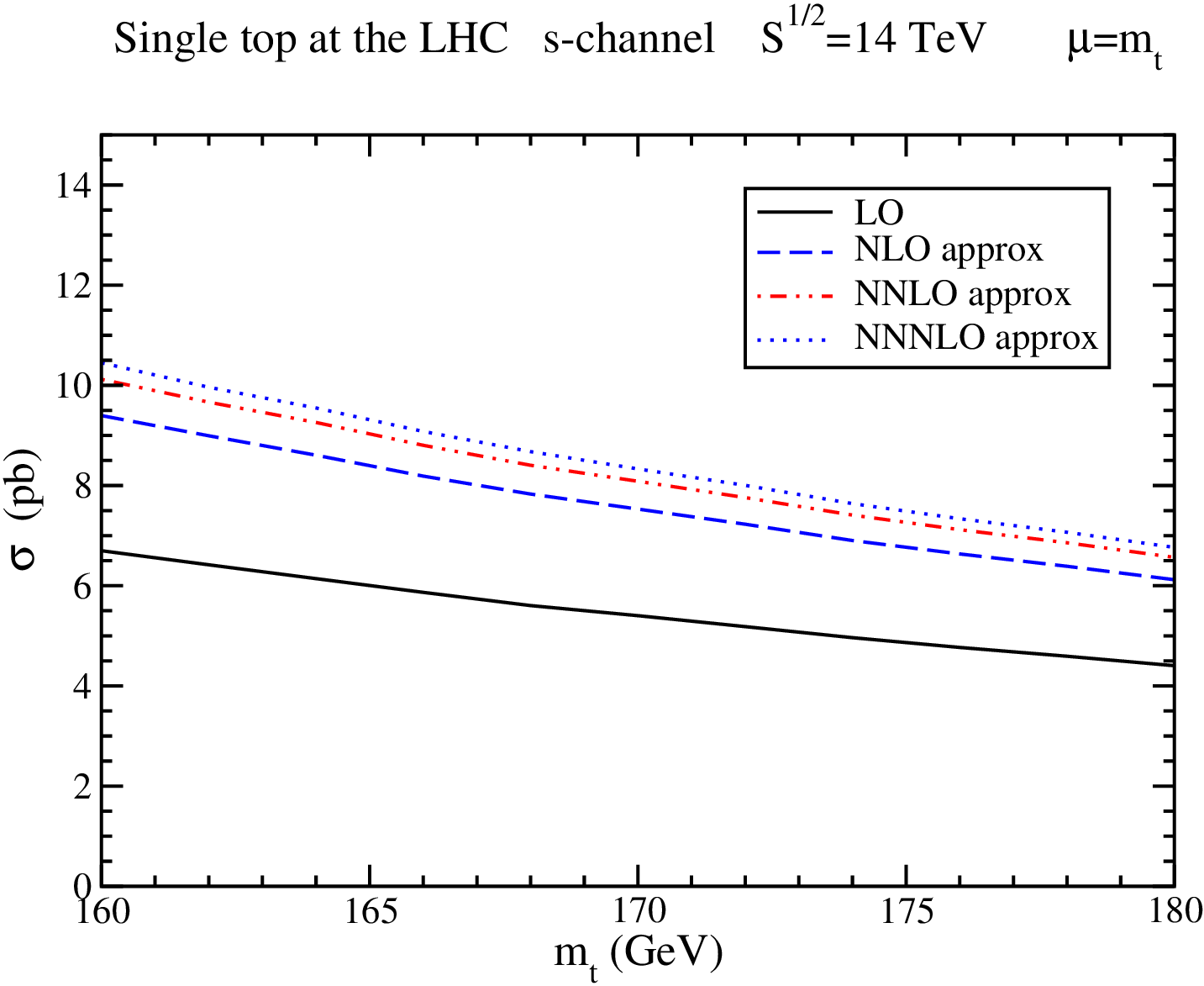}
\hspace{5mm}
\includegraphics[width=6.6cm]{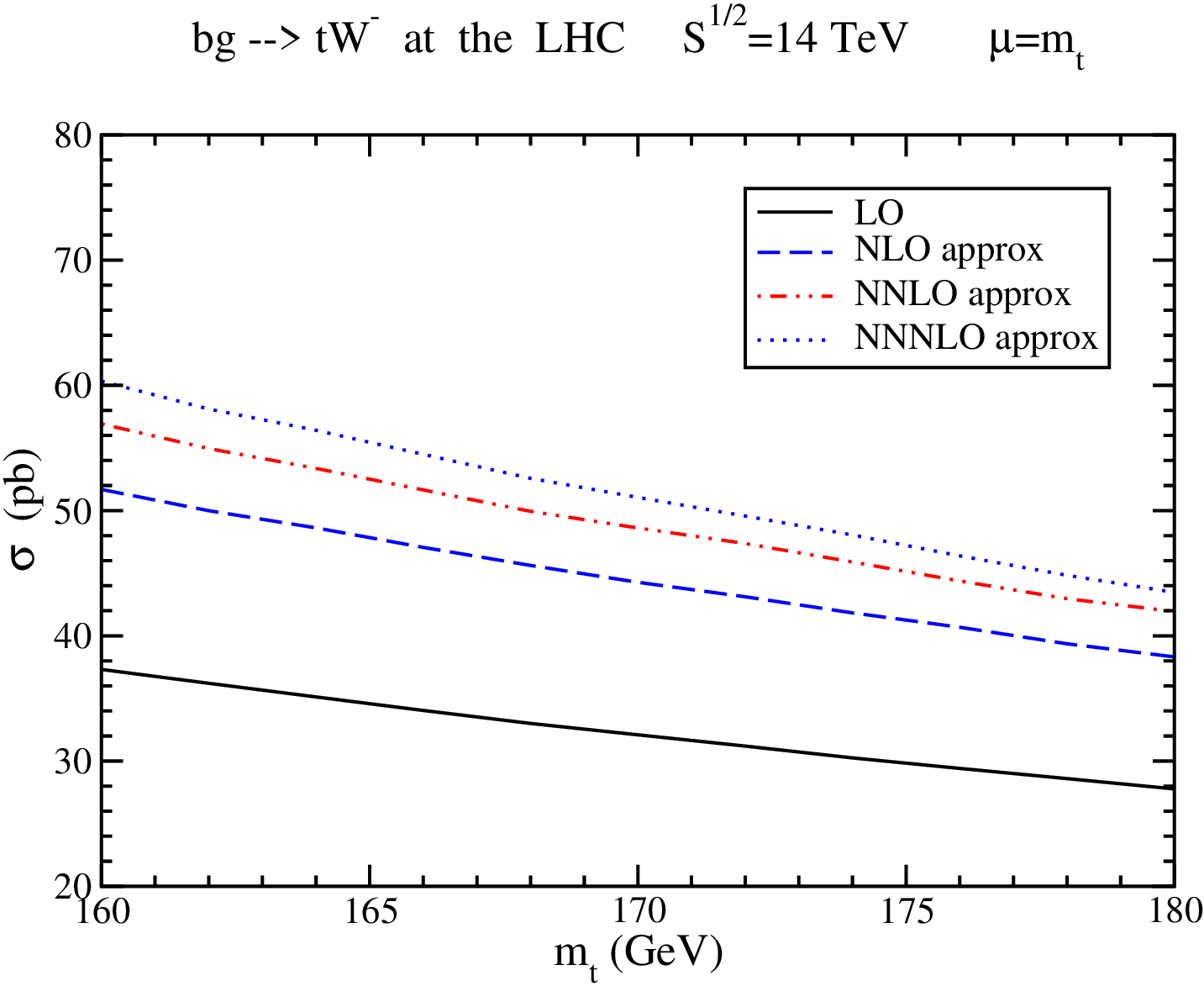}
\caption{$s$-channel single top (left) and $tW$ (right) cross sections 
at the LHC.}
\end{center}
\label{Fig:s-chan.tW.LHC}
\end{figure}

\section*{Acknowledgements}
 
This work has been supported by the National Science Foundation under
Grant No. PHY 0555372.

\end{document}